\documentclass[10pt,twocolumn,english]{revtex4}
\usepackage[T1]{fontenc}
\usepackage[latin9]{inputenc}
\usepackage[a4paper]{geometry}
\usepackage{amsmath}
\usepackage{amssymb}
\usepackage{graphicx}

\makeatletter
\@ifundefined{textcolor}{}
{%
 \definecolor{BLACK}{gray}{0}
 \definecolor{WHITE}{gray}{1}
 \definecolor{RED}{rgb}{1,0,0}
 \definecolor{GREEN}{rgb}{0,1,0}
 \definecolor{BLUE}{rgb}{0,0,1}
 \definecolor{CYAN}{cmyk}{1,0,0,0}
 \definecolor{MAGENTA}{cmyk}{0,1,0,0}
 \definecolor{YELLOW}{cmyk}{0,0,1,0}
}

\makeatother

\usepackage{babel}

\begin{document}

\title{Critical exponents of $O(N)$--models in fractional dimensions}

\author{Alessandro Codello$^a$}
\author{Nicol\'o Defenu$^b$}
\author{Giulio D'Odorico$^c$}

\address{
$^{a}$CP$^{3}$--Origins and Danish IAS, University of Southern Denmark, Campusvej 55, DK-5230 Odense M, Denmark
\\$^{b}$SISSA, Via Bonomea 265, 34136 Trieste, Italy
\\$^{c}$Radboud University Nijmegen, Institute for Mathematics, Astrophysics and Particle Physics,
 Heyendaalseweg 135, 6525 AJ Nijmegen, The Netherlands}

\begin{abstract}
We compute critical exponents of $O(N)$--models in fractional dimensions between two and four, and for continuos values of the number of field components $N$,
in this way completing the RG classification of universality classes for these models.
In $d=2$ the $N$--dependence of the correlation length critical exponent gives us the last piece of information needed to establish a RG derivation of the Mermin--Wagner theorem.
We also report critical exponents for multi-critical universality classes in the cases $N\geq2$ and $N=0$.
%
Finally,  in the large--$N$ limit our critical exponents correctly approach those of the spherical model, allowing us to set $N\sim100$ as threshold for the quantitative validity of leading order large--$N$ estimates.
\end{abstract}

\maketitle

\subsection*{Introduction}

%
The understanding of universality -- namely the independence of the critical properties of a system from its microscopic details --  by means of the renormalization group (RG) has been one emblematic example of the twist of paradigm that such a technique has brought to modern physics.
In Wilson's general framework \cite{Wilson}, the way physics changes with respect to the energy scale is  represented by a flow along a trajectory in a generalized {\it theory space}, which is the space of all theories describing fluctuations of a given set of degrees of freedom.
Critical phenomena arising in a physical system are understood as described by theories that are fixed points of its RG flow \cite{Wilson}.
In this way different trajectories, corresponding to different microscopic theories, which lie in the same basin of attraction of a given fixed point, will describe the same critical properties.
Universality then tells us that these are determined by few parameters, such as the dimensionality, the symmetry group of the system and the order of criticality. Each value of these parameters defines a different universality class; classifying them is tantamount to classifying all possible continuous phase transitions that can occur in Nature.

Among the universal quantities characterising a phase transition, a set of parameters which
acts as a bridge between theory and experiment is that of critical exponents, which parametrize
how certain measurable quantities (such as specific heat, density, susceptibility and so on)
depend on temperature near a critical point. 
Being universal observables, critical exponents are both a test ground for theoretical methods and
possible predictions for, yet unobserved, phase transitions.
Having a simple mathematical tool to compute and predict these exponents is thus an important theoretical and phenomenological task.

%
In this paper we compute the critical exponents of $O(N)$--models in fractional dimensions between two and four and for continuous values of the number of field components $N$, starting from the basic principles of Wilsonian RG in its modern functional realization \cite{Wetterich_1993,Berges_Tetradis_Wetterich_2002}.
$O(N)$--models have many applications to low dimensional systems:
they can describe long polymer chains ($N=0$), liquid--vapor ($N=1$),
superfluid helium ($N=2$), ferromagnetic ($N=3$) and
QCD chiral ($N=4$) phase transitions \cite{Berges_Tetradis_Wetterich_2002,PV}.

%
The present work complements and completes the analysis and classification of universality classes of $O(N)$--models made in \cite{Codello:2012ec} with the dependence of critical exponents $\nu,\alpha,\beta,\gamma,\delta$ on $d$ and $N$, and provides as well the last piece of information needed to give a RG proof the Mermin--Wagner--Hohenberg (MWH) theorem \cite{MWH1,MWH2}.
We also compute the critical exponents for many new $N\geq2$ universality classes describing multi-critical models in fractional dimension $2\leq d \leq3$.
We remark that the critical exponents that we compute are observables for any value of $d$ and $N$ and as such are here first reported.
We also complement the analysis of the possible multi-critical phases of polymeric systems, as found
in our previous work, by giving the critical exponents associated to these phase transitions.
Thus, if these phases can be realised in some system, these can be seen as predictions for parameters yet to be measured.

%
%
%

\subsection*{Scaling solutions and $\eta$}

%

Our tool will be the running effective potential $U_{k}(\rho)$,
which is a function of the $O(N)$--invariant $\rho=\frac{1}{2}\varphi^{2}$,
for a constant field $\varphi$. 
In terms of dimensionless variables $\tilde{U}_{k}(\tilde{\rho})=k^{-d}U_{k}(\rho)$, with $\tilde{\rho}=k^{-(d-2+\eta)}\rho$,
a scaling solution $\partial_{t}\tilde{U}_{*}(\tilde{\rho})=0$ satisfies the following ordinary differential equation \cite{Wetterich_1993}:
%
\begin{eqnarray}
\!\!\!\!\!\!\!\!&&-(d-2+\eta)\tilde{\rho}\,\tilde{U}_{*}'+d\,\tilde{U}_{*} =\nonumber \\
\!\!\!\!\!\!\!\!&& \quad\quad\quad c_{d}(N-1)\frac{1-\frac{\eta}{d+2}}{1+\tilde{U}'_{*}} +c_{d}\frac{1-\frac{\eta}{d+2}}{1+\tilde{U}'_{*}+2\tilde{\rho}\,\tilde{U}_{*}''}\,,
\label{flow}
\end{eqnarray}
%
where $c_{d}^{-1}=(4\pi)^{d/2}\Gamma(d/2+1)$.
The anomalous dimension $\eta$ fixes the scaling properties of the field
at a particular fixed point; to lowest order its value is given
by \cite{Berges_Tetradis_Wetterich_2002}:
$
\eta=4c_{d}\tilde{\rho}_{0}\tilde{U}_{*}''(\tilde{\rho}_{0})^{2}/[1+2\tilde{\rho}_{0}\tilde{U}_{*}''(\tilde{\rho}_{0})]^{2}\,,\label{2}
$
with $\tilde{\rho}_{0}$ the absolute minimum of the fixed point potential $\tilde{U}_{*}'(\tilde{\rho}_{0})=0$.

%
Every solution of (\ref{flow}), together with its domain of attraction, represents a different 
$O(N)$ universality class \cite{Codello:2012ec}.
%
%
%
For every $d$ and $N$ one finds a discrete set of solutions 
corresponding to multi-critical potentials of increasing order, i.e. with $i$ minima,
which describe multi-critical phase transitions
(in which one needs to tune multiple parameters to reach the critical point).
For each of these it is possible to obtain the anomalous dimension $\eta_{i}(d,N)$ (we define $\eta \equiv \eta_2$) as a function of $d$ and $N$,
by means of which we can follow the evolution through theory space 
of the fixed point representing the $i$--th multi-critical potential \cite{Codello:2012sc}.

%
The analysis presented in  \cite{Codello:2012ec} revealed that for $d>4$  and for any $N$,
in accordance with the Ginzburg criterion, one finds only the gaussian fixed point ($i=1$).
(See \cite{Percacci:2014tfa} for a discussion on the possible existence of non--trivial universality classes in $d\geq4$ raised by \cite{Klebanov}).
Starting at $d=4$, the upper critical dimension for $O(N)$--models, the Wilson--Fisher (bi-critical) fixed points ($i=2$) branch away from the gaussian fixed point.
When $d=3$ these fixed points describe the known universality classes of the Ising, {\it XY}, Heisenberg and other models.
%
%
Approaching $d=2$ one clearly observes that only the $N=1$ anomalous dimension continues to grow: for all other values
of $N\geq2$ the anomalous dimension bends downward to become zero exactly when $d=2$.
As explained in \cite{Codello:2012ec}, this non-trivial fact, not evident from the structure of equation (\ref{flow}) alone,
%
is the manifestation of the MWH theorem.
%
%

We now complement this analysis with the results for the correlation length critical exponents $\nu_i(d,N)$ as a function of $d$ and $N$.
We obtained results for the first several multi-critical universality classes $i=2,3,4,5,...$. Here we will only report in detail the analysis for the bi-critical and tri-critical cases ($i=2,3$), and briefly comment on the other multi-critical cases.
%
%
%
%

\subsection*{Eigen--perturbations and  $\nu$}
%
\begin{figure}
\begin{centering}
\includegraphics[scale=0.78]{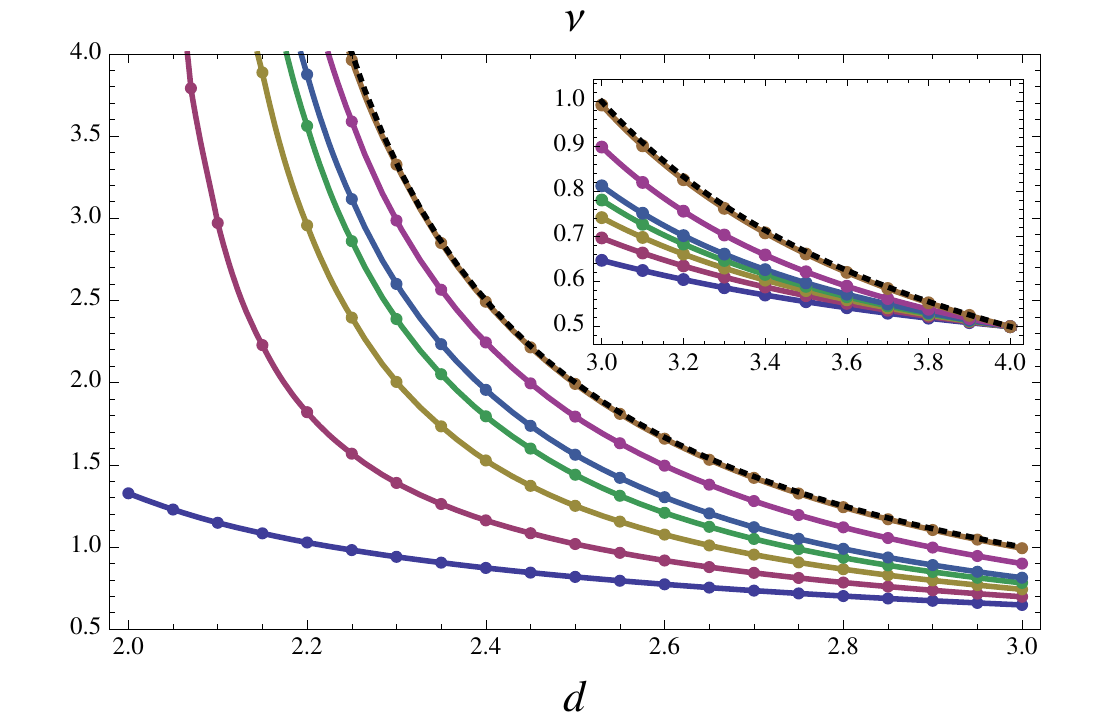}
\par
\end{centering}
\caption{Correlation length critical exponent $\nu$ as a function of $d$ between two and three for $N=1,2,3,4,5,10,100$.
In the inset we show the critical exponent in the range $3\leq d\leq4$.}
\label{nu}
\end{figure}


The correlation length exponent $\nu_i$ is related to the greatest negative (IR repulsive) eigenvalue $y_{1,i}$ of the linearized RG transformation by $\nu_i=1/y_{1,i}$ (we define $\nu \equiv \nu_2$).
In order to calculate it, we will use the eigen--perturbation method described in \cite{Morris97}.
As a starting point, we expand the dimensionless effective potential as follows:
%
$$\tilde{U}_{k}=\tilde{U}_{*}(\tilde{\rho})+\epsilon\,\tilde{u}_k(\tilde{\rho})\,e^{y t}\,,$$
where $\tilde{U}^{*}(\tilde{\rho})$ is a solution of the fixed point equation \eqref{flow} and $\tilde{u}_k(\tilde{\rho})$ is a perturbation around the solution whose eigenvalue is $y$.
Substituting this expression into the flow equation, and considering only terms of first order in $\epsilon$, we obtain an equation for the perturbation:
\begin{eqnarray}
\label{Linear}
&&(d+y)\tilde{u}_k(\tilde{\rho})-(d-2+\eta)\tilde{\rho}\,\tilde{u}_k'(\tilde{\rho})=\nonumber\\
&& -c_{d}(N-1)\left(1-\frac{\eta}{d+2}\right)\frac{\tilde{u}_k'(\tilde{\rho})}{(1+\tilde{U}'_{*}(\tilde{\rho}))^{2}}\nonumber\\
&&-c_{d}\left(1-\frac{\eta}{d+2}\right)\frac{\tilde{u}'_{*}(\tilde{\rho})+2\tilde{\rho}\,\tilde{u}_{*}''(\tilde{\rho})}{(1+\tilde{U}'_{*}(\tilde{\rho})+2\tilde{\rho}\,\tilde{U}_{*}''(\tilde{\rho}))^{2}}\,.
\end{eqnarray}
In order to solve this equation we need two initial conditions. 
The first is obtained by noting that the perturbation equation (\ref{Linear}) is linear, so we can require the normalization condition $\tilde{u}_k(0)=1$, while the second one is imposed on $\tilde{u}_k'(0)$ form continuity at zero field:
$$(y+d)\tilde{u}_k(0)=-c_d\frac{\left(1-\frac{\eta}{d+2}\right)N}{(1+\tilde{U}'_{*}(0))^{2}}\tilde{u}_k'(0)\,.$$
It should be noted that in the special case $N=0$ the continuity at zero field is given by $\tilde{u}_k(0)=0$ and then the normalization condition should be imposed on the first derivative of the perturbation $\tilde{u}_k'(0)=1$.


A generic solution of equation (\ref{Linear}) in the $\rho\rightarrow\infty$ limit behaves at leading order as:
\begin{eqnarray}
\tilde{u}_k(\tilde{\rho}) =a(y)\rho^{\frac{(d-y)}{(d-2+\eta)}}+b(y)e^{C\rho^{\frac{2d}{d-2+\eta}-1}}\,,\nonumber
\end{eqnarray}
where $a(y),\,b(y)$ are two functions of the eigenvalue $y$ and $C$ is a constant depending on $d$ and $\eta$. 
This shows that in the infinite field limit, the solution is a linear combination of power-law and exponential diverging parts \cite{Morris97}.
In order to find the discrete set of eigenvalues that we need, we have to require the solution to grow no faster than a power-law, so the condition is just $b(y)=0$.

\begin{figure}
\begin{centering}
\includegraphics[scale= 0.44]{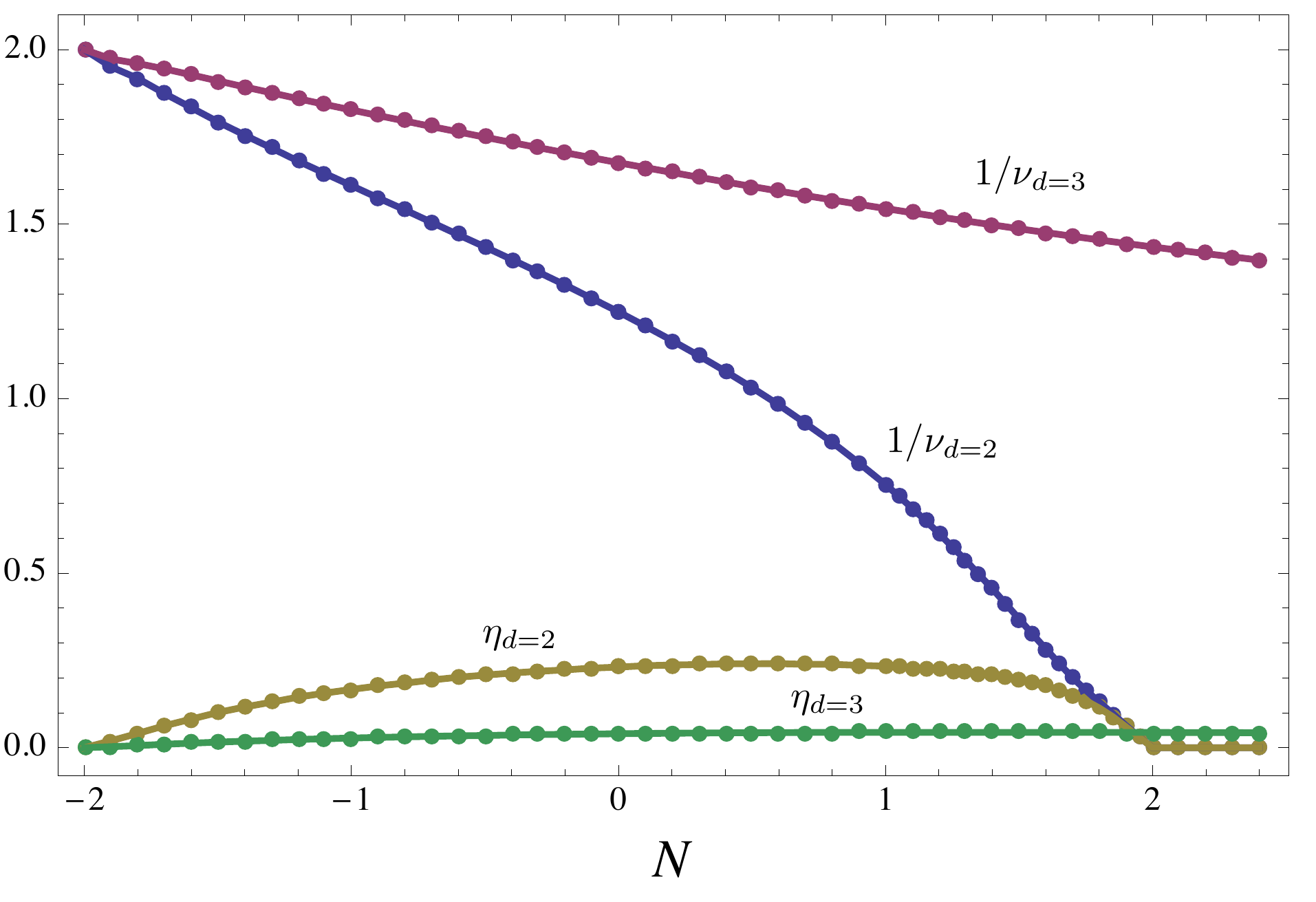}
\par
\end{centering}
\caption{Critical exponents $\eta$ and $y_{1,2}=1/\nu$ as a function of $N$ in two and three dimensions.
The fact that the two dimensional curves are zero for $N\geq2$ is the manifestation of the MWH theorem.
%
}
\label{nuN}
\end{figure}

Using this condition we found just one IR repulsive eigenvalue for the bi-critical fixed point, two for the tri-critical fixed point, three for the tetra-critical fixed point and so on. In this way we were able to construct the curves shown in Figures \ref{nu}, \ref{nuN}, \ref{tri-critical} and \ref{critN}.

The proliferation of eigenvalues is due to the fact that
the $i$--th universality class has $i-1$ IR repulsive directions in theory space, and 
thus we have $i-1$ solutions with negative eigenvalue in the perturbation equation \eqref{Linear}.
In the following we will denote as $y_{j,i}$ the $j$--th eigenvalue of the $i$--th universality class.
%

%
As was already observed in \cite{Codello:2012ec}, the vanishing of the anomalous dimension,
when combined with the behaviour of the $\nu_{i}$ exponents,
implies that there are no continuous phase transitions for $N\geq2$ in $d=2$.
Consistently with this argument, here we find that only the $N=1$ model 
has finite correlation length exponent in two dimensions, in all other cases, $N\geq2$,
$\nu$ blows up as $d\rightarrow2$.
This allows us to distinguish the spherical model, related to the $N\rightarrow\infty$ limit \cite{spherical}, from the gaussian model, both having $\eta=0$.
In the $N\rightarrow\infty$ limit instead we recover the known exact relation $\nu(d,\infty)=\frac{1}{d-2}$ \cite{Morris:1997xj}.

%
Figure \ref{nuN} shows $\eta$ and $y_{1,2}=1/\nu$ as function of $N$ in the interval between $-2\leq N\leq2.5$, for the two cases $d=2$ and $d=3$.
The critical exponents are continuous in the whole range and in particular around $N=0$; this is an indication that the $N\rightarrow0$ limit, relevant to the problem of self avoiding random walks (SAW) \cite{SAW}, is well defined.

These curves represent a strong confirmation of the MWH theorem: for $N\geq2$ both $\eta(2,N)$ and $y_{1,2}(2,N)$ vanish,
while in $d=3$ they have finite values; thus $O(N)$--models with continuous symmetries cannot have a phase transition in two dimensions. 
We remark that both exponents are necessary to distinguish between the case of no phase transition, where we have seen both exponents vanish,
and the $N=\infty$ case where, for example $\eta(3,\infty)$ vanishes but $y_{1,2}(3,\infty)$ attains a finite non--mean field value.
Our computation of $\nu(d,N)$ thus completes the RG derivation of the MWH theorem started in \cite{Codello:2012ec} with the analysis of $\eta(d,N)$.
In the limit $N\rightarrow-2$ both exponents attain their mean field values (namely $\eta=0$ and $\nu=1/2$),
where indeed the model is know to have gaussian critical exponents in both dimensions \cite{Fisher1973}. 

Our functions $\eta(d,1)$ and $\nu(d,1)$ can be compared with results from the bootstrap approach \cite{El-Showk:2013nia}.
The anomalous dimension compares fairly well considering that our computation is based on the solution of a single ODE, while the correlation length critical exponent is slightly overestimated for $d$ in the proximity of two.
It will be interesting to have bootstrap results for the $N>1$ cases in dimension other than three \cite{Kos:2013tga} and in particular to see the emergence of the MWH theorem in this approach.

Finally, to our knowledge, our results are the only one available in the literature regarding the full form of the functions $\eta(d,N)$ and $\nu(d,N)$, functions that are both universal and in principle experimentally accessible for any value of $d$ and $N$.

\begin{figure*}
\includegraphics[scale=0.58]{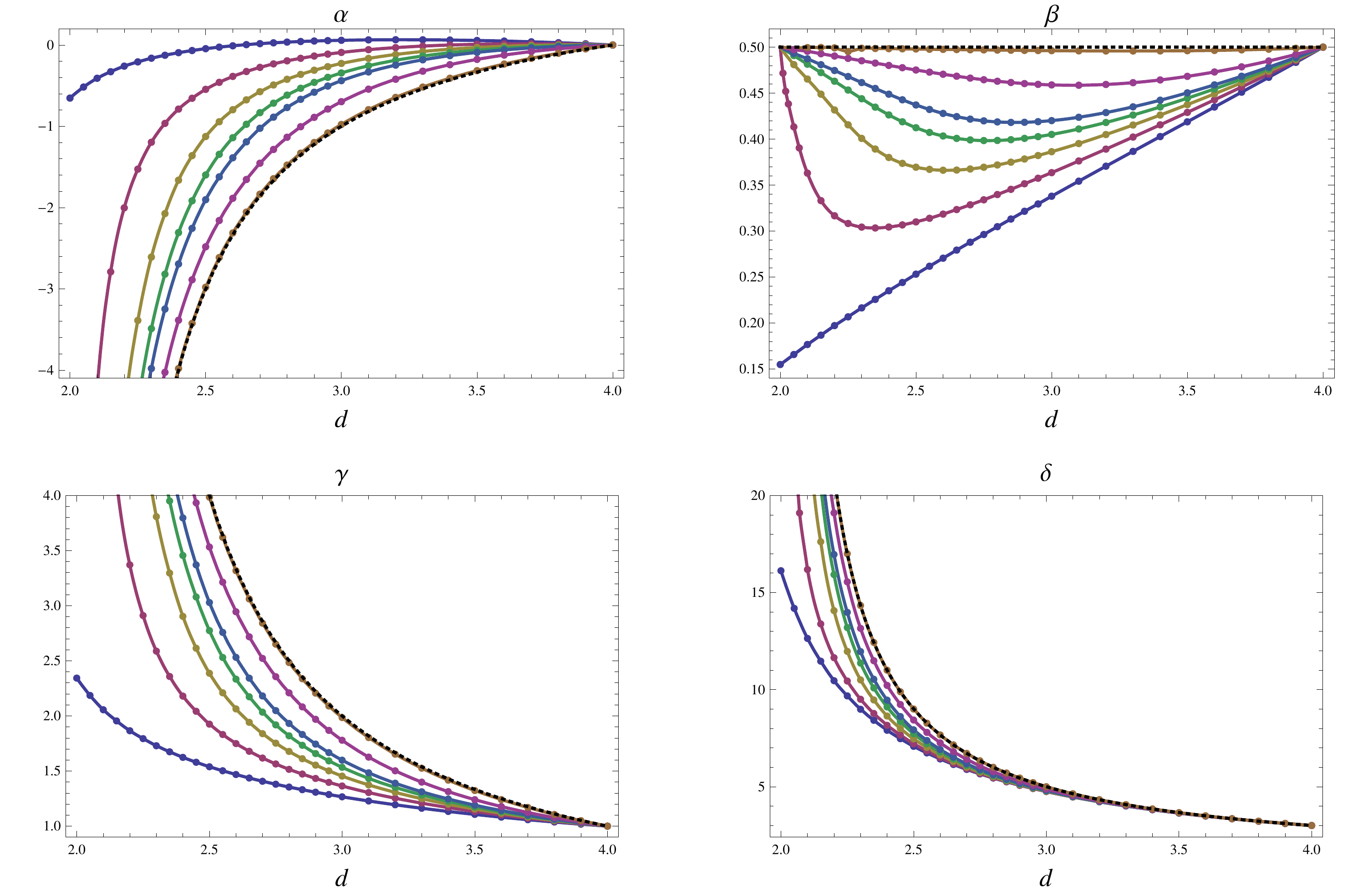}
\caption{
Critical exponents $\alpha, \beta, \gamma, \delta$ of the bi-critical (Wilson--Fisher) universality class for $N=1,2,3,4,5,10,100$.
The $N=1$ curves reach a distinguished value at $d=2$, while the $N =100$ curve is practically equivalent to the exact large--$N$ spherical model limit (represented by dashed lines). The curves for all the other values of $N$ interpolate between the two.}
\label{abcd}
\end{figure*}

\subsection*{Scaling relations and $\alpha,\beta,\gamma,\delta$}

Having obtained $\nu$ and $\eta$ as a function of $d$ and $N$, we can now use the standard scaling relations to obtain the other critical exponents:
\begin{eqnarray}
\alpha &=& 2-\nu d \qquad\qquad \beta	 = \nu\frac{d-2+\eta}{2} \nonumber \\
\gamma &=& \nu(2-\eta) \qquad\qquad  \delta =	 \frac{d+2-\eta}{d-2+\eta} \,.
\label{sr}
\end{eqnarray}
Our results are shown in Figure \ref{abcd} for $2\leq d \leq 4$ and for $N = 1,2,3,4,5,10,100$.

The first thing we notice is that in the large--$N$ limit we smoothly recover the critical exponents of the spherical model \cite{spherical} $\alpha=0$, $\beta=\frac{1}{2}$, $\gamma=\frac{2}{d-2}$ and $\delta=\frac{d+2}{d-2}$. Our results indicate that the $N=100$ case is perfectly approximated by the spherical model, while already at $N=10$ deviations from this limit are appreciable. This shows that, for the regards of computing critical quantities, leading large--$N$ estimates are  quantitatively good only for $N$ of order $10^2$ or larger \cite{Ma}.

For $N=1$ and $d=2$ our results can be compared with the known exact Ising critical exponents found by Onsager \cite{Onsager:1943jn}, which are $\eta^{ex}=0.25$, $\nu^{ex}=1$, $\alpha^{ex}=0$, $\beta^{ex}=0.125$, $\gamma^{ex}=1.75$ and $\delta^{ex}=15$.
Our results are $\nu=1.33$, $\eta=0.23$, $\alpha=-0.65$, $\beta=0.15$, $\gamma=2.34$ and $\delta=16.12$.

Quantitative agreement is not excellent, as expected by the simplicity of our approach, based entirely on the solution of a single ODE (\ref{flow}) and the relative eigenvalue problem (\ref{Linear}).
Still, the insights furnished by our study are very valuable, since they offer a complete qualitative, and almost quantitative, picture of the theory space of $O(N)$--models as a function of both $d$ and $N$.
No other method, to our knowledge, has a similar versatility. In any case, once qualitative understanding has been achieved, one can obtain arbitrarily good quantitative estimates by resorting to higher orders of derivative expansion \cite{Canet_Delamotte_Mouhanna_Vidal_2002}, of which equation (\ref{flow}) represents just the zeroth order.

It is possible to find a better $\nu$ value in the $N=1$ case using a different definition for the anomalous dimension \cite{Codello:2012sc}
rather than the one we used \cite{Codello:2012ec}. This definition, which is strictly valid only in the $N=1$ case, gives a worse value for
$\eta\simeq 0.4$, but a much better result for $\nu=1.01$.

In $d=3$ our $N=1$ results are $\nu=0.65$ and $\eta=0.044$, to be compared with the best known results $\nu=0.63$ and $\eta=0.036$ \cite{kleinert:2001book,ElShowk:2012ht}.
As we see the agreement is much better. This is due to the fact that the derivative expansion can be considered as an expansion in terms of the anomalous dimension: 
the error we commit will then be of the order of the anomalous dimension, which is smaller
in $d=3$ than in $d=2$.
As $N$ grows quantitative estimates become better;
we made comparisons in the cases $N=2,3,4$ and higher and we found good agreement with best known values \cite{kleinert:2001book}.

\subsection*{Tricritical universality class}

\begin{figure}
\begin{centering}
\includegraphics[scale=0.78]{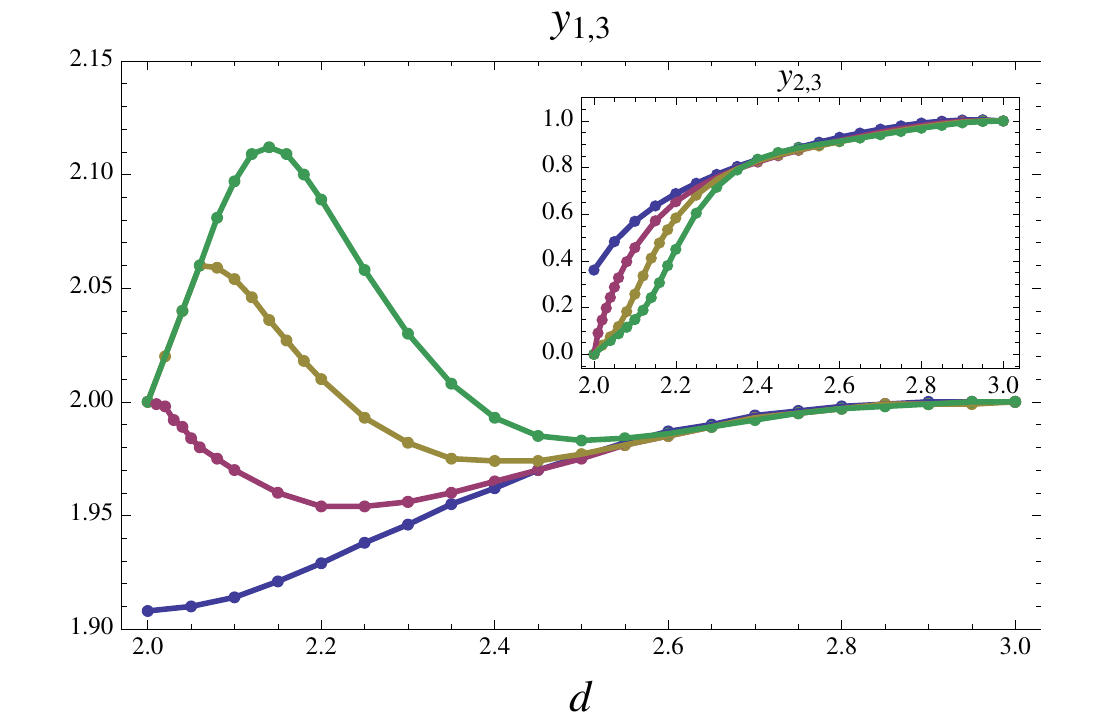}
\par\end{centering}
\caption{Critical exponents $y_{1,3},y_{2,3}$ of the tri-critical fixed points as a function of $d$ for $N=1,2,3,4$. These exponents describe the divergence of the correlation length
as a function of the two critical parameters of the tri-critical universality class.
}
\label{tri-critical}
\end{figure}

In this case we have two IR repulsive eigenvalues of the linearized flow, both shown in Figure \ref{tri-critical}  for $2\leq d \leq 3$, and $N=1,2,3,4$.
The exponent $y_{1,3}=1/\nu_3$ is inverse correlation length exponent; 
indeed at the upper tri-critical dimension, $d_{c,3}=3$, it reaches its mean field value $y_{1, 3}=2$. 
%
When $N=1$ the exponent does not depart
so much from the mean field result as in the standard bi-critical case. 
The $d=2$ value we obtain is $y_{1,3}=1.90$ to be compared with the exact result \cite{Domb:1985jz} $y^{ex}_{1,3}=1.80$, both rather close to the mean field value.
In the case of continuous symmetries ($N\geq 2$) the tri-critical universality class disappears in $d=2$,
and the $y_{1,3}$ exponents correctly returns to their mean field values for every $N$.

The $y_{2,3}$ exponent, instead, describes the divergence of the correlation length as a function
of an additional critical parameter. At the upper tri-critical dimension the mean field
result is $y_{2,3}=1$. When $N=1$ we find in two dimensions $y_{2,3}\simeq 0.4$ which should be compared with
the exact value \cite{Domb:1985jz} $y^{ex}_{2,3} = 0.8$. In this case the agreement is rather low, but this is not surprising since we know that the
LPA$'$ approximation is rather inefficient in $d=2$. However
it should be noted that even if not quantitatively correct, these results can be used to evaluate the crossover exponent
$\phi=\frac{y_{2,3}}{y_{1,3}}$. In $d=2$ this gives $\phi\simeq 0.2$ which, despite the quantitative error, gives a much better estimation
than the $\epsilon$--expansion, which provides a negative value for this exponent at order $\epsilon^{2}$ \cite{Lewis:1978}. 
For continuous symmetries, $N\geq 2$, $y_{2,3}$ vanishes in $d=2$, in the
same way as the exponent $y_{1,2}$ does
in the bi-critical case.

\subsection*{Higher multi-critical universality  classes}

The behaviour of the tri-critical case can be generalized to the other multi-critical universality classes.
For these classes with $i>3$, we have that at the upper critical
dimension, $d_{c,i}=2+\frac{2}{i-1}$ \cite{Codello:2012sc}, all the $i-1$ IR repulsive eigenvalues attain their mean field values.
The largest one will always be $y_{1,i}=2$, as in the standard bi-critical case, 
with all the others having a mean field value smaller than $2$. 
For $N\geq 2$ all the exponents, but the lowest one,
will have different values as a function of $d<d_{c,i}$, all remaining pretty
close to the mean field value, which is eventually recovered in $d=2$. 
Conversely the lowest eigenvalue will decrease monotonically until it vanishes in $d=2$. 
For $N=1$ instead, all the multi-critical universality classes will still exist in $d=2$ and thus all the exponents will reach a finite non--mean field value,
which will be given by the relative CFT result.

\subsection*{The $N = 0$ case}

\begin{figure}
\begin{centering}
\includegraphics[scale=0.78]{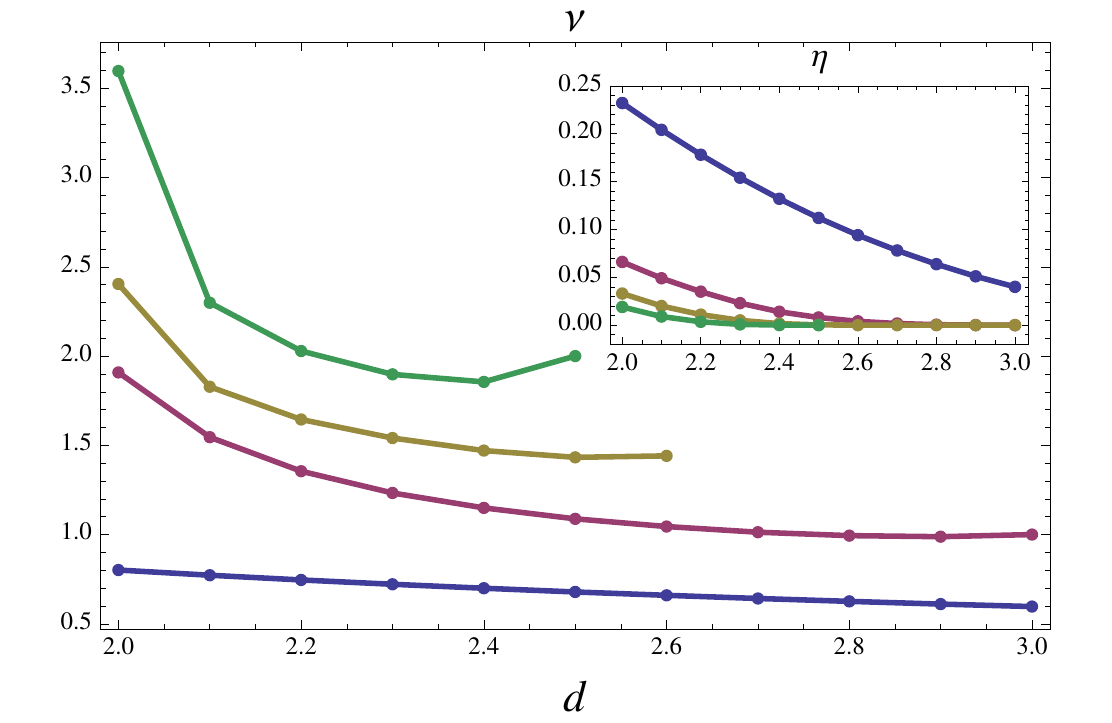}
\par\end{centering}
\caption{Critical exponents in the $N=0$ case. In the main plot are shown the values of $\nu_i$ in the range $2\leq d \leq 3$ for the (from the bottom) bi-critical, tri-critical, tetra-critical and penta-critical universality classes, corresponding respectively to $i=2,3,4,5$. In the inset the corresponding values of $\eta_i$ are reported (in inverted order, from top to bottom).}
\label{critN}
\end{figure}

Multi-critical scaling solutions are also found for $N=0$, which survive in infinite number
when $d\rightarrow2$ \cite{Codello:2012ec}.
A plot of $\eta_i$ and $\nu_i$ for the first four universality classes $i=2,3,4,5$ is shown in Figure \ref{critN}; 
these are numerically very similar to those of the $N=1$ cases
(see \cite{Codello:2012ec} and Figure \ref{nu} for the bi-critical class).
This was indeed expected, judging from Figure \ref{nuN}.


In $d=2$ the exact results for (bi-critical) SAW \cite{Nienhuis}, which correspond to the $N = 0$ limit of $O(N)$--models \cite{SAW}, together with scaling relations give:
$\eta^{ex}=5/24\simeq 0.208$, 
$\nu^{ex}=3/4 = 0.75$, 
$\alpha^{ex}=0.5 $, 
$\beta^{ex}=5/64\simeq 0.078$, 
$\gamma^{ex}=43/32\simeq 1.344$, 
$\delta^{ex}=91/5\simeq 18.2$. 
We find a good agreement: 
$\eta=0.232$, 
$\nu=0.801$, 
$\alpha=0.398$, 
$\beta=0.093$, 
$\gamma=1.416$, 
$\delta=16.24$.
In $d=3$ we can compare with the Monte Carlo results \cite{PV}: 
$\eta^{MC}= 0.028$, 
$\nu^{MC}= 0.587$, 
$\alpha^{MC}= 0.239$, 
$\beta^{MC}= 0.302$, 
$\gamma^{MC}= 1.157$, 
$\delta^{MC}= 4.837$. 
We find again a reasonably good agreement:
$\eta=0.04$, 
$\nu\ =0.597$, 
$\alpha =0.210$, 
$\beta =0.310$, 
$\gamma =1.169$, 
$\delta =4.769$.
From these comparisons we see that the $N=0$ estimates are better than the $N=1$ estimates, since also the $N\geq2$ estimates are so, this indicates that the (bi-critical) Wilson--Fisher universality class is the one for which our estimates are poorer.

We are not aware of any known result regarding multi-critical phase transitions of polymeric systems, or any other model that belongs to one of the $N=0$ multi-critical universality class.
Our estimates for the critical exponents are given in Figure \ref{critN} and to our knowledge these result are novel predictions: it will be interesting to find physical systems or theoretical models described by these universality classes to test them.
\\

\subsection*{Conclusions}

In this paper we reported the computation of critical exponents of $O(N)$ universality classes as a function of the dimension and of the number of field components.
The correlation length critical exponent $\nu$ was computed by studying the eigenvalue problem obtained 
linearizing the RG flow of the running effective potential around the scaling solutions found in \cite{Codello:2012ec}, representing the $O(N)$ multi-critical	fixed point theories.
From this and the previous knowledge of the anomalous dimensions,
all the remaining exponents $\alpha$, $\beta$, $\gamma$, $\delta$ were found using scaling relations.

In particular we displayed the critical exponents for the bi-critical (Wilson--Fisher) and tri-critical phase transitions for general $d$ and $N$.
Another result which is new to our knowledge are the critical exponents for the multi-critical classes in the $N\to 0$ limit. These, via the De Gennes correspondence \cite{SAW}, are universal, observable quantities which can be associated to possible new phases of polymeric systems.
To the best of our knowledge, this physics is yet to be observed.

One interesting feature which is worth mentioning is that there is a correspondence between critical exponents of models with short--range interactions in fractional dimension and models with long--range interactions in integer dimension \cite{Defenu:2014bea}.
This means that our curves $\eta(d,N)$ and $\nu(d,N)$ have direct physical interpretation, not only for systems in fractional dimensions, but as describing the critical behaviour of models with long--range interactions in two or three dimensions. In this case our universal results could be indirectly tested in the near future, both by numerical simulations and laboratory experiments. Further details on this correspondence can be found in \cite{Defenu:2014bea}.

By computing the function $\nu(d,N)$ we provided the information necessary to complete the RG proof of the Mermin--Wagner--Hohenberg theorem, as put forward in our previous work \cite{Codello:2012ec}. This constitutes a first important example of how one can use RG equations to give precise statements on how universality classes depend on dimension and symmetry group parameters, a general and fundamental problem whose solution has important applications in physical model building in both condensed matter and high energy physics.

We conclude by stressing that here we explored just the simplest realization of our method and this alone allowed a complete qualitative understanding of $O(N)$ universality classes. We believe that its numerical results, where not fully satisfactory, can be fairly improved in future extensions along the lines explained in the text, and will ultimately lead to a definitive quantitative understanding of critical properties of $O(N)$--models.\\
\\
\textit{Acknowledgements.} 
The CP$^3$-Origins centre is partially funded by the 
Danish National Research Foundation, grant number DNRF90.


\end{document}